\documentclass[preprint2]{emulateapj}

\shorttitle{Astrometry}
\shortauthors{Guyon et al.}
\usepackage{graphicx}
\usepackage[mathscr]{eucal}
\begin{document}

\title{SIMULTANEOUS EXOPLANET CHARACTERIZATION AND DEEP WIDE-FIELD IMAGING WITH A DIFFRACTIVE PUPIL TELESCOPE}

\author{Olivier Guyon}
\affil{Steward Observatory, University of Arizona, Tucson, AZ 85721, USA}
\affil{National Astronomical Observatory of Japan, Subaru Telescope, Hilo, HI 96720, USA}
\email{guyon@naoj.org}
\author{Josh A. Eisner, Roger Angel, Neville J. Woolf}
\affil{Steward Observatory, University of Arizona, Tucson, AZ 85721, USA}
\author{Eduardo A. Bendek, Thomas D. Milster}
\affil{College of Optical Sciences, University of Arizona, Tucson, AZ 85721, USA}
\author{S. Mark Ammons}
\affil{Lawrence Livermore National Laboratory, Physics Division L-210, 7000 East Avenue Livermore, CA 94550, USA}
\author{Michael Shao, Stuart Shaklan, Marie Levine, Bijan Nemati}
\affil{Jet Propulsion Laboratory, 4800 Oak Grove Drive, Pasadena, CA 91109, USA}
\author{Frantz Martinache}
\affil{National Astronomical Observatory of Japan, Subaru Telescope, Hilo, HI 96720, USA}
\author{Joe Pitman}
\affil{Exploration Sciences, PO Box 24, Pine, CO 80470, USA}
\author{Robert A. Woodruff}
\affil{Lockheed Martin, 2081 Evergreen Avenue, Boulder, CO 80304, USA}
\author{Ruslan Belikov}
\affil{NASA Ames Research Center, Moffett Field, CA 94035, USA}

\begin{abstract}
High-precision astrometry can identify exoplanets and measure their orbits and masses, while coronagraphic imaging enables detailed characterization of their physical properties and atmospheric compositions through spectroscopy. In a previous paper, we showed that a diffractive pupil telescope (DPT) in space can enable sub-$\mu$as accuracy astrometric measurements from wide-field images by creating faint but sharp diffraction spikes around the bright target star. The DPT allows simultaneous astrometric measurement and coronagraphic imaging, and we discuss and quantify in this paper the scientific benefits of this combination for exoplanet science investigations: identification of exoplanets with increased sensitivity and robustness, and ability to measure planetary masses to high accuracy. We show how using both measurements to identify planets and measure their masses offers greater sensitivity and provides more reliable measurements than possible with separate missions, and therefore results in a large gain in mission efficiency. The combined measurements reliably identify potentially habitable planets in multiple systems with a few observations, while astrometry or imaging alone would require many measurements over a long time baseline. In addition, the combined measurement allows direct determination of stellar masses to percent-level accuracy, using planets as test particles. We also show that the DPT maintains the full sensitivity of the telescope for deep wide-field imaging, and is therefore compatible with simultaneous scientific observations unrelated to exoplanets. We conclude that astrometry, coronagraphy, and deep wide-field imaging can be performed simultaneously on a single telescope without significant negative impact on the performance of any of the three techniques.
\end{abstract}

\section{Introduction}
\label{sec:intro}

Among the existing techniques to identify exoplanets, astrometry and direct imaging are particularly well suited for identification and characterization of nearby habitable planets. Either technique can provide a full census of habitable planets around nearby (approximately $d < 10$ pc) F,G,K main-sequence stars, provided that it achieves the required sensitivity (sub-$\mu$as single measurement precision for astrometry, $10^{-9}$ raw contrast at $\approx 100 $mas for coronagraphy). Extensive characterization of the exoplanet does however require both techniques, which provide complementary information.
\begin{itemize}
\item{Direct imaging with a high-contrast instrument is required to acquire spectra allowing characterization of the planet's atmosphere. Direct imaging also reveals the exoplanet's environment (other planets in the system, structure of the exozodiacal cloud), and allows measurement of the planet's rotation period and weather variability through time-series photometry.}
\item{Astrometry\footnote{In this paper, absolute astrometry (measurement of the absolute position of a star on the sky) is referred to as the {\it astrometry} measurement, while relative astrometry (measurement of the relative offset between an exoplanet and its host star, usually from a coronagraphic image) is referred to as the {\it coronagraphic} measurement.} is a promising technique to measure the mass of nearby exoplanets down to a fraction of an Earth mass, and this measurement is required to gain a physical understanding of the planet's surface and atmosphere. Although radial velocity (RV) measurements can also measure planet masses, their ability to reach Earth-mass planets around Sun-like stars is limited by stellar jitter (we however note that an Earth-mass planet in the habitable zone of a low-mass star can produce a signal that exceeds stellar jitter).}
\end{itemize}

A planet's ability to retain an atmosphere and the composition of this atmosphere depend strongly on its mass. For low-mass planets, light observed by direct imaging may for example originate from the planet's surface, while planets with mass larger than a few Earth mass will likely retain a dense and opaque atmosphere hiding it. Meaningful interpretation of a planet's photometry and color/spectrum will therefore require prior knowledge of its mass, as colors alone can be ambiguous \citep[see, for example,][]{2010ApJ...724..189C}.

\begin{figure*}
\includegraphics[scale=0.45]{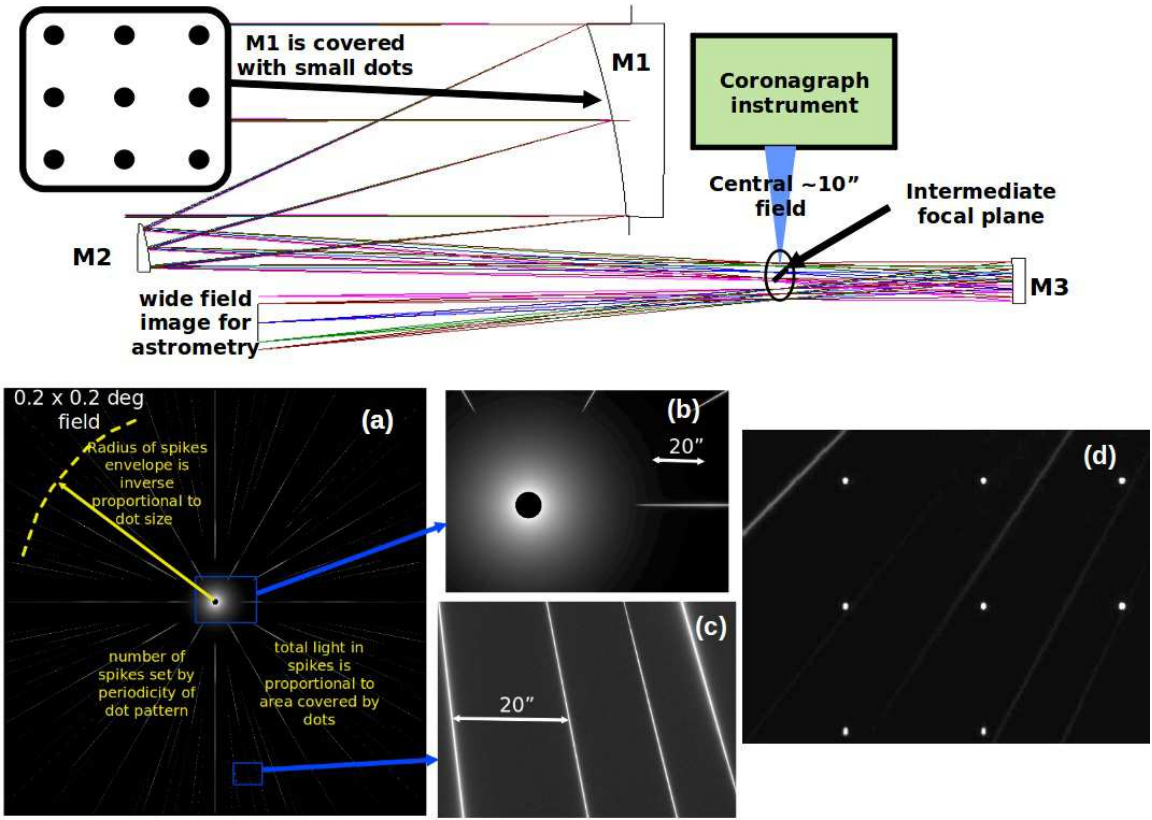} 
\caption{\label{fig:optprinciple} Conceptual optical design for the diffractive pupil telescope (DTP) proposed by \cite{DPTpaper1}. The top part of this figure shows how light is shared between two instruments. The central field containing the bright star and its immediate surroundings is extracted at the telescope's intermediate focus and fed to a coronagraph instrument for high-contrast imaging. The wide well-corrected outer field is imaged onto a large focal plane detector array. Panels (a)-(d) show details of the wide-field image acquired in the final focal plane. (a) The wide-field image shows the diffraction spikes introduced by the dots on the primary mirror. (b) The central part of the target image, containing most of the flux, is missing from the wide-field image as it has been directed to the coronagraph instrument. (c) Faint diffraction spikes pave the rest of the field. (d) Faint background stars are imaged simultaneously with the diffraction spikes. While images (a)-(c) are simulated, image (d) was acquired in a laboratory demonstration of the technique \citep{2011SPIE.8151E..26B}.}
\end{figure*}

Several astrometry \citep{2005ASPC..338...37U,2009astro2010S.271S,2011ExA...tmp...87M} and direct imaging concepts \citep{2009arXiv0911.3200L,2009SPIE.7440E..13G,2010ASPC..430..375T,2010ASPC..430..368S,2006SPIE.6265E..38C} have been proposed and studied for identification and characterization of nearby exoplanets, demonstrating the technical feasibility of either measurement approach. While the scientific value of both astrometry and direct imaging for characterization of exoplanets is widely recognized, either path requires a dedicated mission with its own technological challenges. Obtaining both astrometric and direct imaging measurements was until recently assumed to require two separate missions, and simultaneous development, construction and operation of both missions would likely be beyond available resources. 

\cite{DPTpaper1} recently proposed using a diffractive pupil telescope (DPT) to obtain both types of measurements simultaneously. The astrometric concept relies on simultaneous imaging of numerous background stars and diffraction spikes from the brighter target star on the same focal plane detector array. The diffraction spikes are created by a regular grid of dark spots on the telescope primary mirror. The submillimiter diameter dots cover a few percent of the primary mirror area, and their regular arrangement in the pupil plane creates in monochromatic light a regular grid of diffraction-limited spots in the focal plane. In broadband light, the spots are stretched into long thin spikes radiating from the central bright star (the target). Astrometric distortions in the imaging system (including the detector) are accurately measured by tracking the position of the spikes on the detector array, enabling high-precision astrometric calibration without requiring picometer-level stability of the telescope optics or focal plane detector array. The astrometric measurement is performed by differential position measurements of the diffraction spikes originating from the bright central star targeted for planet search, relative to field stars. Because the spikes are generated by the primary mirror, such measurements are largely immune from astrometric distortions, as they equally affect the spikes and the field stars. The unique feature of DPT is that it allows simultaneous coronagraphic imaging within a narrow-field instrument along with astrometry and general astrophysics imaging with a wide-field camera. An optical design for the DPT is shown in Figure \ref{fig:optprinciple} for a 1.4 m diameter telescope. This baseline configuration, adopted by \cite{DPTpaper1}, offers a 0.2 $\mu$as single-axis single-measurement astrometric precision for a 0.3 deg$^2$ field of view. This precision is a steep function of the telescope diameter, and improves, to a lesser extent, with the camera's field of view.

While our first paper \citep{DPTpaper1} was focused on the technical aspects of the DPT concept and its expected astrometric precision and accuracy, we discuss and quantify in this paper its ability to identify and characterize exoplanets and perform deep wide-field observations that may not be related to exoplanet science. 
In Section \ref{sec:exoplanetmass}, we quantify the required astrometric measurement precision for mass determination of exoplanets detectable by a coronagraph, and illustrate the scientific benefits of simultaneous coronagraphic imaging and astrometric measurement. We discuss in Section \ref{sec:planetarysystems} the observation of planetary systems, which may consist of several planets. Finally we show in Section \ref{sec:deepim} that the DPT concept allows deep wide-field imaging at the full sensitivity offered by the telescope, and that the impact of the added diffraction spikes on wide-field imaging sensitivity is negligible, even when the telescope is pointed at a bright star during exoplanet science observations, suggesting that the telescope could also be valuable for general astrophysics purposes. 
Detection and mass measurement of exoplanets with astrometry alone, as was previously considered for astrometry mission concepts, is briefly discussed in the Appendix and referred to when discussing the scientific merit of the proposed technique.

\section{Exoplanet orbital parameters and Mass measurement -- Single-planet case}
\label{sec:exoplanetmass}

\subsection{Science Goal and Representative Example}

The science goals of the astrometric measurement explored in this paper are twofold: 
\begin{itemize}
\item{Assist the coronagraph to detect exoplanets. At the minimum, the astrometric measurement should confirm detections performed by the coronagraph and help constrain the orbital parameters of all planets identified by coronagraphic imaging. It may also uncover planets which are too close to the coronagraph's performance limits (especially inner working angle, IWA) to be firmly identified from coronagraph images alone, even if the planet appears in at least one of the coronagraph observations.}
\item{Measure the mass of all planets imaged by the coronagraph in the habitable zone of nearby stars}
\end{itemize}
In this section, we quantify the astrometric measurement precision required to meet these goals. We note that the second goal is more challenging than the first one, and that the performance required can therefore be derived from the second goal (mass measurement) alone.

We assume that a 1.4 m telescope is used with a coronagraph offering a 2 $\lambda$/D IWA, following the baseline design adopted by \cite{DPTpaper1}, and inspired from the Pupil mapping Exoplanet Coronagraph Observer mission concept \citep{2010SPIE.7731E..68G}. We consider that the number and duration of observations is driven by the coronagraph instrument's goal to identify and acquire spectra of potentially habitable planets around a few high-priority targets: each observation is 48 hr long, and each high-priority target is observed approximately every two months. The expected astrometric precision of this system has been described in \cite{DPTpaper1}, and was found to be 0.2 $\mu$as per axis per measurement for a 0.3 deg$^2$ field of view camera. We note that the astrometric precision is independent of the coronagraph instrument design and performance, as it is achieved with an optical path which is separate from the coronagraph instrument. 

The relative mass estimation precision achieved for a fixed instrument design is a function of the planet type, stellar brightness (bright stars are easier, as their spikes are brighter in the astrometric camera), and star distance (the astrometric signal is smaller for more distant stars). 
We choose to adopt a Sun analog at 6 pc from Earth as a representative example of a challenging target for detection of an Earth-like planet with the coronagraph. The star apparent magnitude is $m_V$=3.7. We choose to place an Earth-mass planet at 1.2 AU around this star with a 1.3 rad inclination, therefore avoiding complications associated with the 1 yr period blind spot in the astrometric measurement. We use this example, for which the planetary system characteristics are listed in the top part of Table \ref{tab:obsmodel}, to quantify the relationship between astrometric measurement precision and mass estimation precision.

\begin{deluxetable*}{lc}
\tablecolumns{2} 
\tablewidth{0pc} 
\tablecaption{\label{tab:obsmodel}Observation Model} 
\tablehead{ 
\colhead{Parameter} & \colhead{Value}}

\startdata 
\multicolumn{2}{c}{{\bf Planetary system characteristics}}\\
Star                 		& Sun analog\\
Distance             		& 6 pc \\
Location            		& Ecliptic pole \\
Orbit semi-major axis 		& 1.2 AU \\
Orbital period       		& 1.3 yr \\
Planet mass          		& 1 Earth mass\\
Orbit inclination               & 1.3 rad\\
Orbit eccentricity         	& 0.2 \\
Astrometric signal amplitude 	& 0.5 $\mu$as\\
Orbit apparent semi-major axis  & 200 mas\\
\multicolumn{2}{c}{{\bf Measurements}}\\
Number of observations          & 32 (regularly spaced every 57 days)\\
Coronagraphic image: planet position precision & 2.5 mas per axis (=3.6 mas in 2D) \tablenotemark{a}\\
Coronagraphic image: inner working angle (IWA) & 130 mas\\
Astrometry: single-measurement precision     & Variable (driven by science requirement)\\
\enddata 
\tablenotetext{a}{Corresponds to an S/N=10 detection with $\lambda/D = 80 $mas (single-axis astrometric precision is theoretically equal to $(\lambda/D)/(\pi \sqrt{N_{ph}})$). For a photon-noise-limited measurement with no background, this would be achieved with $N_{ph}=$100 photons at 550nm for a 1.4m telescope.}
\end{deluxetable*}

\subsection{Planet Mass Measurement}
\label{ssec:planetmassmeas}

In this section, we describe a simple model of simultaneous observation of the system by the coronagraph and the astrometric camera. We then use the model to estimate the astrometric precision required to achieve the science goals outlined in the previous section.

{\it Measured quantities}. The planet orbit and mass are measured by combining the simulated astrometric and coronagraphic data. Coronagraphic and astrometric observations are assumed to be simultaneous. For $N$ epochs, the total set of measurements consists of 4$N$ values: 2$N$ absolute astrometric measurements and 2$N$ relative planet to star astrometric measurements derived from the coronagraphic images. Unless otherwise noted, simulations for this paper use a total number of 32 epochs, regularly spaced every 57 days to span a total of 5 yr. With each observation assumed to be 2 day long, 28 targets could be observed at this cadence.
 
\begin{figure}
\includegraphics[scale=0.6]{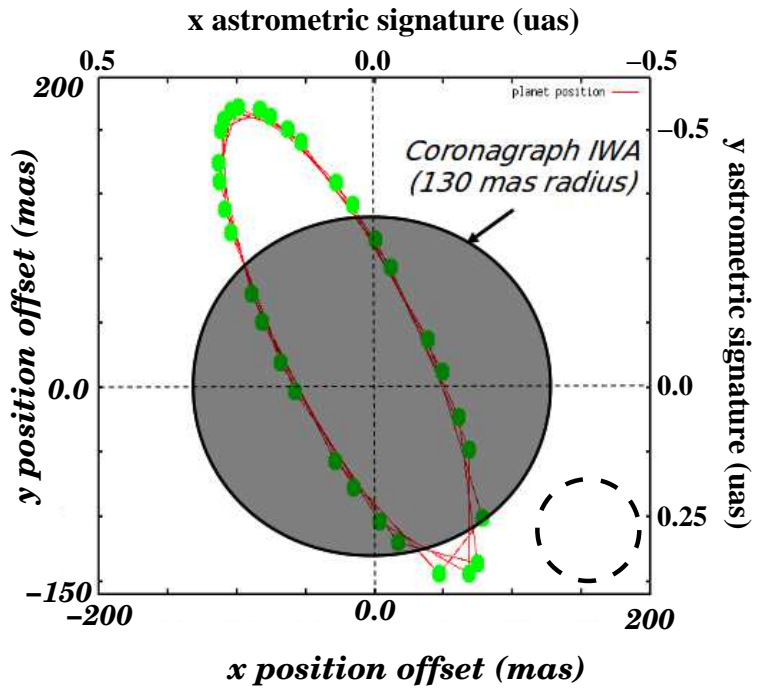} 
\caption{\label{fig:planetorbit} 
Coronagraphic (relative astrometric position of the planet referenced to the star) and astrometric (position of the star referenced to the sky) signatures are opposite in sign and of very different amplitude. The scale for the coronagraphic measurement is shown in the bottom and left, while the scale for the astrometric measurement is shown at the top and right. The dark shaded disk shows the zone within which the planet cannot be seen by the coronagraph. The dashed circle diameter is the one-dimensional standard deviation of a single astrometric measurement (0.2 $\mu$as). The standard deviation of the coronagraph measurement (2.5 mas) is about half the size of the small green dots showing the planet position along its orbit.}
\end{figure}

{\it Planet image position measurements from coronagraphic images}. We assume that each coronagraph observation yields the apparent position of the planet relative to the star with a 2.5 mas 1$\sigma$ precision per axis, provided that the planet is outside the inner working of the coronagraph (assumed to be 130 mas). The assumed 2.5 mas 1$\sigma$ precision per axis is equal to 1/30 $\lambda /D$ at $\lambda = 500 \mu$m, and would be achieved with an signal-to-noise ratio (S/N)=10 detection, or, in the photon-noise limit without background, with approximately 100 photons. While in a 2 day exposure, the number of photons collected from the planet is larger than 100, other sources of error will limit the measurement accuracy, such as centering error of the star on the coronagraph mask, and effect of uncorrected stellar speckles on the planet photocenter estimation. We have somewhat arbitrarily adopted 1/30 $\lambda /D$ as the single-axis measurement precision in this study, noting that a more detailed evaluation would be required to establish this measurement precision with high confidence. We note that centering errors in the coronagraph can be calibrated by introducing faint ghosts of the central star, as proposed by \cite{2006ApJ...647..620S} and \cite{2006ApJ...647..612M}, and such a scheme is likely to be required to reach the 1/30 $\lambda /D$ accuracy assumed here. If the planet is within the coronagraph mask, the position measurement error is increased to the size of the focal plane mask, essentially removing the contribution of this measurement from the overall solution. The apparent position of the planet on the sky is shown in Figure \ref{fig:planetorbit} for each of the 32 observations. The planet can only be seen by the coronagraph for 17 out of the 32 observations, as it is located within the coronagraph's IWA for 15 of the observations.

{\it Known variables}. The following three parameters are assumed to be known: 
\begin{itemize}
\item{{\it Star apparent location on the sky (two variables)}. The parallax motion is therefore perfectly known if the star distance is known, and the aberration of light is therefore assumed to be perfectly known as it is only a function of source position on the sky}
\item{{\it Average radial velocity}. The perspective acceleration effect produced by the relative radial motion of the star is therefore assumed to be known, and is not included in our simulations. For a star at 6pc with a 20 km.s$^{-1}$ velocity, the perspective acceleration ranges from 0 to 0.2 mas.yr$^{-2}$ depending on the perspective angle (angle between three-dimensional velocity motion of the target and observer), and is maximum for a 45$^{\circ}$ perspective angle. If unknown, perspective acceleration will only affect the detection of planets with period longer than the observation time span, for which there is a degeneracy between the perspective acceleration and the exoplanet astrometric signature. For planets with orbital period comparable or less than the observation time span, the constant astrometric acceleration term is not strongly coupled with the planet signal, and can be included as a free parameter to the astrometric data fit with little impact on planet detection and characterization performance. The degeneracy described above may be solved with rRV measurement of the host star to directly measure the three-dimensional velocity and thus calibrate the perspective acceleration. With a 1 m.s$-1$ RV measurement accuracy, and proper motion measurement obtained by astrometry with a 0.02 $\mu$as/yr accuracy as shown in Table \ref{tab:combsolution}, analytical derivations show that the error on the perspective acceleration is 0.02 $\mu$as.yr$^{-2}$ (this error is maximum when the perspective angle is close to 90$^{\circ}$, corresponding to a near zero RV). Over a 5 yr mission duration, the cumulative error around the mid-point observation reaches 0.06 $\mu$as at the beginning and end points of the observation series, and is therefore comparable to the astrometric accuracy assumed in this paper. An RV estimate with poorer accuracy is not very useful, as the perspective acceleration is then better constrained by astrometry measurements alone. In this work, we do not take perspective acceleration into account in the global fit, implicitly assuming that (1) stellar RV is known to within 1 m.s$-1$ or (2) the observation time span is comparable or larger than the planet orbital period. Astrometric recovery of long-period planets (which is not the focus of this paper) would require a more careful treatment of perspective acceleration, and may be challenging as these planets are unlikely to be visible in coronagraphic images (reflected light decreases rapidly with orbital radius).}
\end{itemize}
In addition, it is assumed that observing epochs and standard deviation of all measurements (for both astrometry and coronagraphy) are known.

{\it Stellar mass}. We assume that stellar mass is known to a 5\% 1-$\sigma$ accuracy prior to the observations, but we also treat it as a parameter to be solved for. This is done in our model by treating stellar mass as a free parameter and adding, separately from astrometric and coronagraphic measurement, a single simulated stellar mass measurement with a 5\% 1$\sigma$ accuracy.

{\it Correlation between measurements, systematic errors}. We assume that all measurement errors are uncorrelated and all measurements are free of systematic errors. The validity of this assumption for the astrometric measurements is discussed in our previous publication on the DPT concept \citep{DPTpaper1}, and based on the fact that the astrometric measurement is inherently differential. Existing performance estimates reflect our current assessment that dominant error sources in this system will be random, and that concurrent errors occurring in the different astrometric and coronagraphic fields of view will be primarily uncorrelated.  However, we have begun to expand this assessment using detailed modeling and simulation, supported by lab testing, to characterize key error sources and quantify allocations of those errors in a comprehensive system error budget.  We are particularly careful to identify any systemic error sources effecting the entire observational system, as well as the degree of correlation possible between errors both within our coronagraph and between our two concurrent observations.

{\it Free parameters to be solved for}. A planetary model is constructed and linked to the observed quantities. The model is defined by 11 free parameters: central star distance (one variable), proper motion (two variables). and mass (one variable), planet mass (one variable), and orbital parameters (six variables). The values of the 11 free parameters are derived from the maximum likelihood solution given the noisy measurements listed above.

Uncertainties for all 11 free parameters are determined using a Monte Carlo (MC) approach. For the single-planet case explored here, a total of 1000 simulations was used. Each data set is generated assuming a normal distribution, centered on the perfect solution, with standard deviation equal to the respective uncertainties of coronagraphy and astrometry.


\begin{deluxetable*}{lccc}
\tablecolumns{4} 
\tablewidth{0pc} 
\tablecaption{\label{tab:combsolution} Uncertainties in the Combined (Astrometry and Coronagraphy) and Separate (Astrometry or Coronagraphy) Solutions\tablenotemark{a}} 
\tablehead{ 
\colhead{\bf Parameter} & \multicolumn{3}{c}{{\bf 1-$\sigma$ uncertainty}}\\
\colhead{} & \colhead{Astrometry} & \colhead{Astrometry} & \colhead{Coronagraphy}\\
\colhead{} & \colhead{Only} & \colhead{+Coronagraphy} & \colhead{only\tablenotemark{b}}}

\startdata 
parallax                		& 0.037 $\mu$as       & 0.035 $\mu$as      & 2.949 $\mu$as \\
x proper motion             		& 0.017 $\mu$as.yr$^{-1}$    & 0.012 $\mu$as.yr$^{-1}$   & 1.304 $\mu$as.yr$^{-1}$ \\
y proper motion           		& 0.020 $\mu$as.yr$^{-1}$   & 0.013 $\mu$as.yr$^{-1}$   & 1.288 $\mu$as.yr$^{-1}$ \\
Planet mass 				& 0.132 $M_{Earth}$   & 0.098 $M_{Earth}$  & 5.355 $M_{Earth}$\\
Semi-major axis (SMA)      		& 0.0228 AU           & 0.0052 AU          & 0.0047 AU\\
Orbital phase          			& 0.653 rad           & 0.039 rad          & 0.039 rad\\
Orbit inclination         		& 0.0968 rad          & 0.0065 rad         & 0.0060 rad\\
Position angle of SMA on sky 		& 0.111 rad           & 0.0040 rad         & 0.0039 rad\\
Orbit ellipticity  			& 0.098               & 0.0035             & 0.0034\\
Position angle of perihelion          	& 0.648 rad           & 0.0034 rad         & 0.0033 rad\\
Stellar mass 	\tablenotemark{c}	& 0.05 $M_{Sun}$      & 0.013 $M_{Sun}$    & 0.012 $M_{Sun}$\\
\enddata 
\tablenotetext{a}{Assumed measurement precisions: 0.2 $\mu$as per axis per measurement for absolute astrometry, 2.5 mas per axis per measurement for relative astrometry derived from coronagraphic images.}
\tablenotetext{b}{Assumes that astrometry is available at the 20$\mu$as per axis per observation from another mission to constrain parallax. Astrometric measurement at this precision level ($\gg \mu$as) only affects estimates of parallax and proper motion, and has no significant effect on other measured parameters.}
\tablenotetext{c}{Assumed to be known to 5\% accuracy independently of astrometry and coronagraphy measurements}
\end{deluxetable*}

\subsection{Single-observation Astrometric Precision Required to Meet Science Goals}

\begin{figure}
\includegraphics[scale=0.35]{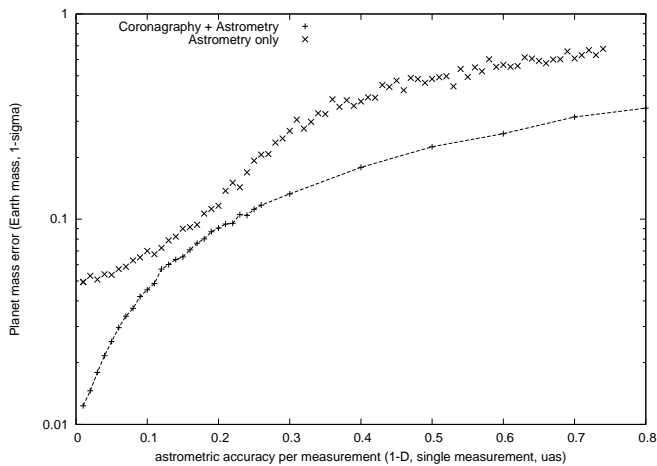} 
\caption{\label{fig:mplaneterr} 
Error on planet mass measurement as a function of single-measurement astrometric precision. The lower curve and points show the planet mass measurement precision when both coronagraphic and astrometric data are used to constrain the solution. When only astrometric data are used, the mass measurement is less accurate. Each point in this figure shows the standard deviation of a set of 1000 solutions, each computed from an independent set of simulated data. 
}
\end{figure}

To estimate the astrometric measurement precision required to meet science requirements, simultaneous coronagraphic and astrometric measurements were simulated and fitted according to the scheme described in Section \ref{ssec:planetmassmeas}. For each simulation, the single astrometric measurement precision was chosen in the range between 0 (perfect astrometry) and 0.8 $\mu$as while the coronagraphic performance (both IWA and precision of the planet image position measurement) were kept constant.
Figure \ref{fig:mplaneterr} shows how the planet mass estimate changes as a function of the level of astrometric error per measurement, and shows that a 10\% relative precision in the estimate of the Earth-mass planet requires a 0.2 $\mu$as precision per astrometric measurement. 

The figure also shows that the mass estimate derived from the combined astrometry and coronagraphy measurements is more accurate than can be obtained from the astrometric measurements only. The difference between astrometry alone and astrometry + coronagraphy is especially significant when high-precision mass measurement is to be obtained (left part of Figure \ref{fig:mplaneterr}), as inferring the planet's mass from astrometry alone is then constrained by uncertainties in the stellar mass, assumed to be at the 5\% level in this example. In this regime, improving the astrometric precision below 0.1 $\mu$as does not significantly improve the planet mass measurement. With both astrometry and coronagraphy, the stellar mass is also measured, and high-precision planet mass measurement is possible. When the astrometric precision is relatively poor (right part of the figure) the astrometry+coronagraphy measurements yield a planet mass estimate standard deviation which is approximately half as large as would be obtained with astrometry alone. For example, with a 0.8 $\mu$as precision per measurement, the planet mass is still constrained $\pm 0.35 M_{Earth}$ with the combined measurements, while astrometry only measurements would not be sufficient to unambiguously detect the planet and estimate its mass. The astrometry only points in the figure are consistent with the findings of the double blind study summarized in the Appendix: for the $0.8 \mu$as precision per astrometric measurement, the mission astrometric S/N is $\approx 4.5$, insufficient (mission S/N $<$ 6) to unambiguously detect the Earth-like planet.
This difference in mass measurement precision, due to the propagation of orbital parameters errors to the mass estimate, is described in the next section. Interestingly, the relative difference in planet mass error between the two scenarios is the smallest when the single measurement astrometric precision is around 0.2 $\mu$as, which is the case considered in this paper for numerical evaluations.

\begin{figure*}
\begin{center}
\includegraphics[scale=0.55]{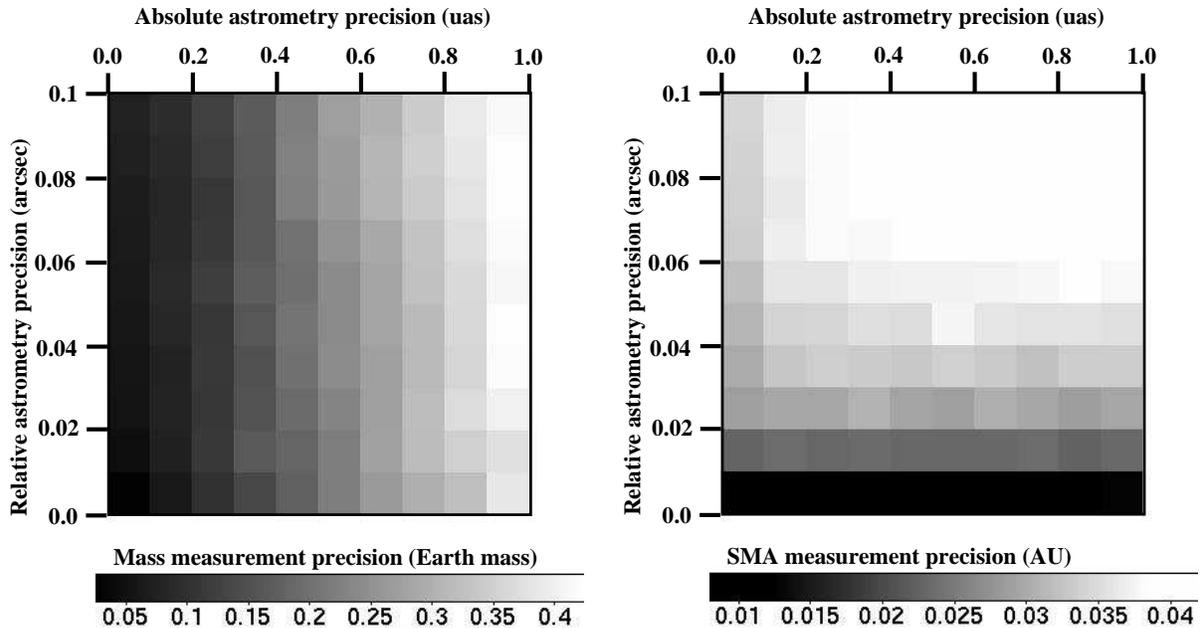} 
\caption{\label{fig:esterrmap} 
Error on planet mass measurement (left) and planet orbit semi major axis (right) as a function of single-measurement relative (vertical axis) and absolute (horizontal axis) astrometric precision. The maps contain white noise due to the finite number of simulated measurements used to produce the figure. 
}
\end{center}
\end{figure*}

Figure \ref{fig:esterrmap} shows how the precision on the planet mass and semi major axis (SMA) varies as a function of both the relative and absolute astrometry measurement precisions. While the precision of the planet mass estimation (left) is mostly a function of absolute astrometry measurement precision, the precision of planet's SMA estimate is mainly a function of relative astrometric precision. In both cases, both measurements however participate to the parameter estimation. This figure illustrates the complementarity between the two measurements.

\subsection{How Coronagraphic Imaging Helps Astrometry to Measure Planet Mass}

While coronagraphic images (providing relative astrometry between the star and the plan) alone do not constrain planet mass, they constrain the values for the orbital parameters much more precisely than possible with absolute astrometric measurements alone. The relative astrometric signal given by the direct image has a large S/N advantage over absolute astrometry: for a 1 $M_{Earth}$ planet on a 200 mas radius orbit, the assumed 2.5 mas uncertainty corresponds to 1/80th of the orbit radius. Constraining the orbital separation to the same level with astrometry alone would require a 0.006 $\mu$as astrometric precision (1/80th of the matching astrometric amplitude).

Solving for planet mass using the combined astrometry and coronagraphy measurements is therefore very powerful, and allows more accurate mass determination than would be possible with astrometry alone.
\begin{itemize}
\item{The coronagraphic images constrain the orbital parameters and reduce error propagation from orbital parameter to mass estimate. For instance, \cite{1999PASP..111..169T} demonstrates how even a single direct image (providing an instant measurement of the projected orbital separation) dramatically improves the constraint on the companion mass in binary systems for which only partial astrometric measurements are available.}
\item{With both coronagraphic imaging and astrometry, both the stellar and planet masses are directly measured. Relative astrometry provides the total mass, while the absolute astrometric signal is driven by the mass ratio: both are required for accurate measurement of the planet's mass.}
\end{itemize}

The 1-$\sigma$ uncertainty for all parameters of the fit is shown in Table \ref{tab:combsolution}, for three cases: astrometric measurement only, combined astrometric and coronagraphic measurements, and coronagraphic measurement alone (assuming a lower precision astrometric measurement is available to constrain the parallax). Adding coronagraphic images to the astrometric data reduces the standard deviation on orbital parameters by approximately a factor 10. The stellar mass is also directly measured, while with astrometry only, it is assumed to be known with a 5\% standard deviation. The planet mass is estimated with a standard deviation below 0.1 $M_{Earth}$ with the combined data, while it would be 35\% larger with astrometry only. Table \ref{tab:combsolution} shows that the uncertainties in planet orbit geometry and stellar mass are identical (to the numerical accuracy of our test) in the astrometry+coronagraphy and coronagraphy only cases, illustrating that these parameters are mostly constrained by the coronagraphic imaging data.

\subsection{Improving Mass Measurement Precision for Planets with a Nearly 1 yr Period}
The astrometric signature of a planet in a 1 yr period orbit can be partially absorbed in the parallax fitting of the astrometry measurements. With astrometric measurements only, the mass estimate error therefore grows as the planet period becomes closer to 1yr for particular orbit geometries. The problem is fundamentally that with finite S/N ratio the two signatures (parallax motion and planet orbital motion) cannot be separated if their frequencies are close. The width of this blind spot is thus reduced as the astrometric measurements span a longer period, corresponding to a better resolution in the frequency domain. For a 5 yr mission duration, the astrometric blind spot covers approximately 5\% of the habitable zone, so the detection rate is still good (approximately 95\%) but is not 100\%. 
The relative astrometry obtained from the coronagraphic images is not affected by parallax motion, and therefore yields an estimate of the orbital parameters that does not have any ambiguity with parallax motion. With coronagraphic and astrometric data, the blind spot is therefore greatly reduced in size (although still not completely eliminated, as planet mass is still ambiguous if the planet orbit and phase matches exactly the parallax motion), especially for a short-duration mission. This benefit is critical for a combined astrometry + coronagraphy mission, as confirmation that a planet is of sufficiently low mass to be potentially habitable needs to occur sufficiently early to allow allocation of significant coronagraphic follow-up time (for spectroscopy and time monitoring).

To illustrate how coronagraphic images mitigate the astrometric blind spot problem, we consider an Earth-mass planet at 1.01 AU from a Sun mass star (period = 1.015 yr) at the ecliptic pole, orbiting a star at 6pc. We assume circular orbits for both the Earth and the target planet, and a planet orbit phase equal to Earth orbit phase plus 1 rad, with a face-on orbit. This particularly unfavorable configuration is used as an example to illustrate the blind spot mitigation thanks to coronagraphy, but is not a representative case of a habitable planet. The system is observed 32 times over 5 yr with observations regularly spaced in time. To demonstrate the robustness of the combined solution, we assume here measurement precisions that are poorer than in the previous section: standard deviation of 0.3 $\mu$as and 5 mas per measurement per axis for respectively astrometry and coronagraphy (instead of 0.2 $\mu$m and 2.5 mas, respectively). With only the astrometry measurements, the model described above yields a mass estimate with a standard deviation of 4.17 $M_{Earth}$: the planet is not detected. With both the astrometric and coronagraphic imaging, the planet mass is estimated with a standard deviation of 0.16 $M_{Earth}$.

\subsection{Discussion}
\label{ssec:singleplanetdiscussion}

We have so far evaluated the scientific value of combining astrometry and coronagraphy by considering a single example (5 yr observation of an Earth-like planet around a Sun-like star), and we now briefly discuss the applicability of the results presented in this section to other observation scenarios. While the orbit geometry (viewing angle, orbit eccentricity) are not expected to significantly impact detection sensitivity, the planet location (semi-major axis), planet and stellar masses, system distance and observation time span must be considered to generalize the findings presented so far.

The astrometric measurement is most sensitive to outer planets (the signal grows linearly with planet separation and planet mass) provided that the observation time span is not significantly shorter than the orbital period. The coronagraphic observations are most sensitive to inner planets (reflected light signal proportional to the inverse square of planet separation) provided that the separation is larger than the coronagraph's IWA, and the measurement sensitivity is only weakly function of planet mass (the reflected light is proportional to the planet's squared radius, which goes as the 1/3rd power of mass for low-mass planets, and increases less rapidly with mass for larger planets). 
The example considered in this section shows that combined astrometric + coronagraphic measurement can identify inner planets near the coronagraph's IWA and measure their mass to approximately 0.1 Earth-mass accuracy. As the planet is moved outward, the astrometric sensitivity increases, allowing mass estimation to a $\approx$ 0.12/SMA Earth mass, where SMA is the orbit semi major axis in AU. At a few AUs separation at most, the planet however becomes too faint to be imaged, regardless of its mass. The optimal mission duration is therefore close to the period of the outermost planets that can be imaged with the coronagraph, and there is little benefit in extending the mission duration further.

Both astrometry and coronagraphy are more sensitive to potentially habitable planets as the stellar mass is reduced (stronger astrometric signal, shorter period, and more favorable reflected light contrast). The reduced planet's angular separation however becomes challenging for the coronagraph, and the performance of the combined measurement scheme becomes driven by the coronagraph's IWA.

As the system's distance to Earth is increased, both the astrometry and coronagraphic imaging signals become weaker, but this effect is much steeper for the coronagraphic imaging than for the astrometric measurement. The relative astrometric position derived from images suffers from the combined effects of smaller flux (planet flux scales as the inverse square of distance) and the angular resolution in absolute unit (scales linearly with distance). The planet is also more likely to fall within the coronagraph's IWA, and the relative contribution of zodiacal and exozodiacal backgrounds increases

\section{Benefit of combining astrometry and direct imaging for characterization of planetary systems}
\label{sec:planetarysystems}

\subsection{Enhancement of Spectral Characterization Efficiency with Astrometry}

The benefits of astrometry for an exoplanet characterization mission have been previously studied under the assumption that an astrometric measurement would occur prior to a direct imaging mission \citep{2010ApJ...720..357S,2011PASP..123..923D,2009SPIE.7440E...9S}, and are twofold.
\begin{itemize}
\item{First, astrometric measurements would reveal the existence of planets and measure their masses, and therefore identify scientifically valuable targets for a future direct imaging mission aimed at acquiring spectra.}
\item{Second, the orbital parameters derived from the astrometric observations could allow optimal timing of the direct imaging observations, when the planet is near maximum elongation. Telescope time can therefore be used more efficiently, as an imaging only mission can be relatively inefficient to find planets, as quantified by \cite{2005ApJ...624.1010B}.}
\end{itemize}
\cite{2010ApJ...720..357S} show that an astrometric precursor mission can reduce the coronagraphic observation time by a factor 2--5 thanks to these two advantages. The last advantage (ability to optimally time direct imaging observations) is however largely lost if several years separate the two missions, as the propagation of orbital parameters error over a long time span does not allow reliable prediction of the planet's apparent position. 

The first advantage (identification of scientifically interesting targets) is immune from error propagation of orbital parameters, and would allow considerable time saving, as the direct imaging mission would spend less time searching for planets and more time characterizing them. A direct imaging mission alone is not efficient for identifying planets which spend a small fraction of their time outside the coronagraph's IWA, and the planet yield may be relatively small in that case. This is an unfortunate situation, as for most habitable planets the angular separation will statistically be close to the coronagraph's IWA. An imaging-only mission aimed at characterizing potentially habitable planets may also devote a significant fraction of time acquiring spectra of planets outside the mass range suitable for habitability, as apparent brightness is a poor proxy for mass. With prior astrometric measurements, these risks are mitigated, and only a few direct imaging observations are required to recover the phase of the planet. Reconnecting the astrometric measurements with the direct imaging detection then allows accurate derivation of the orbital parameters if the time span between the measurements is large.

In this paper, we propose to perform the astrometric and direct imaging measurement simultaneously, so we must reconsider the benefit of astrometry for improving the mission efficiency. Assuming a fixed mission duration, identification of high-priority targets must be performed rapidly in order to allow sufficient time in the second part of the mission for extended characterization. An optimal observing strategy for a multi-year mission would be to first perform high-cadence, short-duration observations to identify potentially habitable planets, followed by several years of follow-up in order to characterize the planet(s) through at least one full orbital period. Rapid identification of planets is especially challenging in multiple planets systems due to confusion between the astrometric signatures.
In the following subsection, we use simulated observations of a multiple planets system to test how rapidly high-priority targets could be identified.

\subsection{Multiple Planets}

We have simulated observations of a planetary system consisting of three planets around a Sun-like star at 6pc. The mass, SMA and orbital period for each of the planets are given in Table \ref{tab:planetsyst}.

\begin{deluxetable}{lcccc}
\tablecolumns{4} 
\tablewidth{0pc} 
\tablecaption{\label{tab:planetsyst} Multiple Planets System Characteristics} 
\tablehead{ 
\colhead{} & \colhead{Mass} & \colhead{SMA} & \colhead{Period} & \colhead{Inclination}\\
\colhead{} & \colhead{($M_{Earth}$)} & \colhead{(AU)} & \colhead{(yr)} & \colhead{(sin($i$))}}

\startdata 
Planet 1    & 1.0 & 1.2 & 1.31 & 0.25\\
Planet 2    & 4.0 & 1.8 & 2.41 & 0.25\\
Planet 3    & 16.0 & 2.4 & 3.72 & 0.25
\enddata 
\end{deluxetable}
 
Telescope and instrument design are identical to what was used in the previous simulations. Following the precision requirement defined in Section \ref{sec:exoplanetmass}, we assume a 0.2 $\mu$as uncertainty per astrometric measurement per axis. 

The goal of this section is to evaluate the combined measurement's scientific value when observing a realistic planetary system. We therefore include several planets (three planets). While the astrometric precision is independent of the planetary system complexity, the relative astrometric measurement derived from coronagraphic images will be affected by confusion issues (such as the potential inability to link planets to their images for widely separated observations, relative astrometric error due to unknown planet affecting the planet image photocenter measurement, and spatial structure in the exozodiacal cloud). For simplicity, we have chosen to not include these effects in our model, but to arbitrarily account for them by adopting a poorer relative astrometry precision. The relative astrometry measurement precision is thus assumed to be 0.0088 arcsec (1/10th of the diffraction limit at $\lambda = 550$ nm) per axis per measurement for each of the three planets, provided that the planet image is outside the coronagraph's IWA.

To illustrate the complementarity of astrometric and coronagraphic measurements, and the value of performing both measurements, we have simulated two different mission scenarios.
\begin{itemize}
\item{{\it Astrometric mission}. One observation is performed every month, with an 0.2 $\mu$as precision per axis.}
\item{{\it Astrometry + coronagraphy mission}. Every two months, an astrometric measurement at the 0.2 $\mu$as precision per axis is performed, simultaneously with a coronagraphic observation. The planet positions are measured to 0.0088 arcsec precision per axis  only when outside the coronagraph's IWA.}
\end{itemize}

\begin{figure*}	
    \begin{center}
   \begin{tabular}{c}
  \includegraphics[width=150mm]{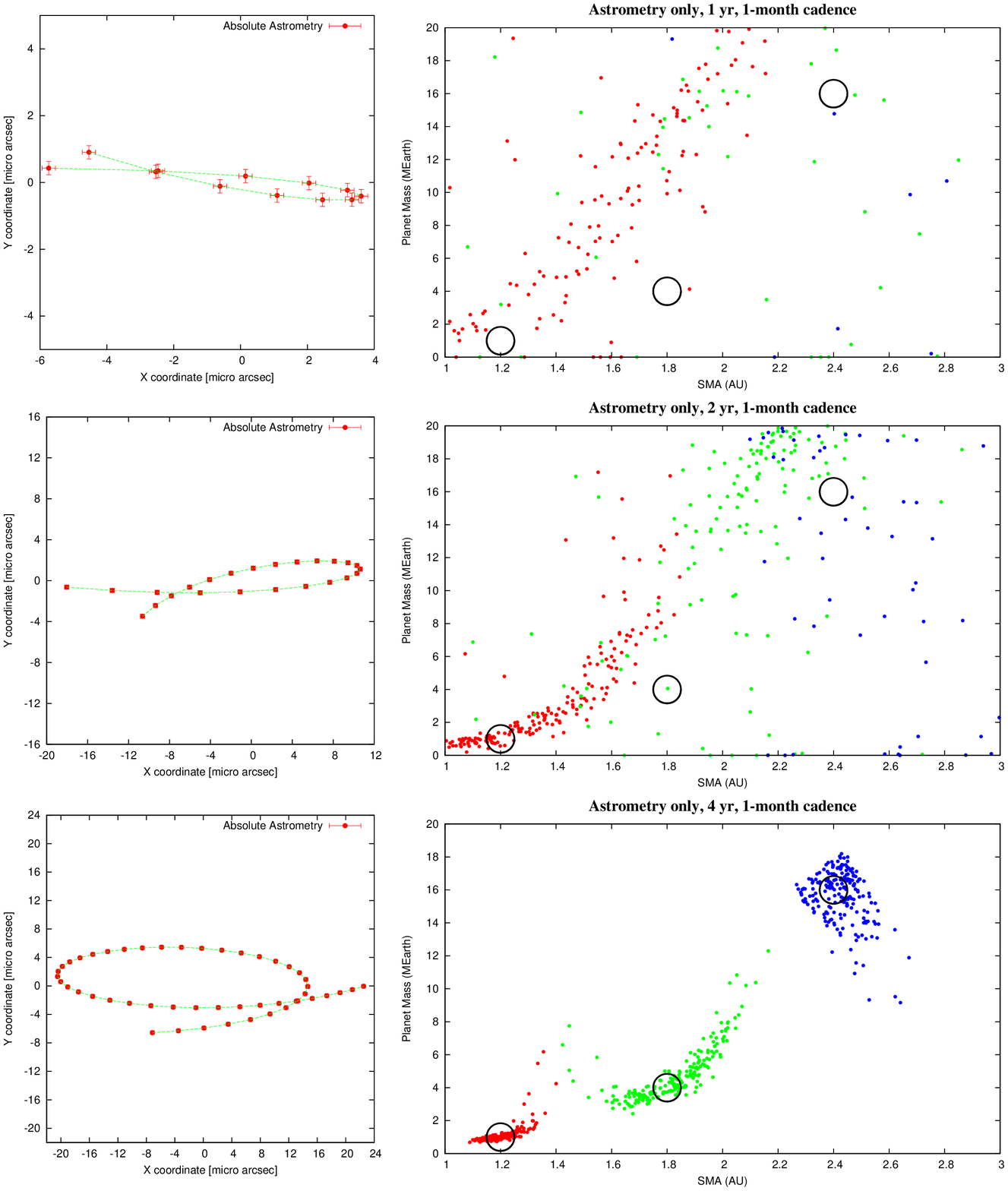} 
   \end{tabular}
   \end{center}
   \caption[multiplanetsol] 
   { \label{fig:multiplanetsolA} 
Result of the astrometry mission scenario, observing a three planet system around a Sun-like star at 6pc. Values to be measured are shown on the left with measurement error bars (1$\sigma$). Two hundred realizations of the observations are simulated, and the best three-planet solution is shown on the right for each simulation in the semi major axis--mass plane. Total mission duration is increased from top to bottom, from 1 yr to 4 yr. The three dark circles indicate the actual configuration for the three-planet system. The initial starting point is the same for all scenarios: the first six absolute astrometry values are identical in all cases. However, the astrometric values shown in the left panel are corrected for uniform drift (proper motion) during the measurement period, resulting in different shapes and amplitude for the part of the astrometry trajectory curves common to all scenarios (first year).}
   \end{figure*} 

\begin{figure*}	
    \begin{center}
   \begin{tabular}{c}
  \includegraphics[width=150mm]{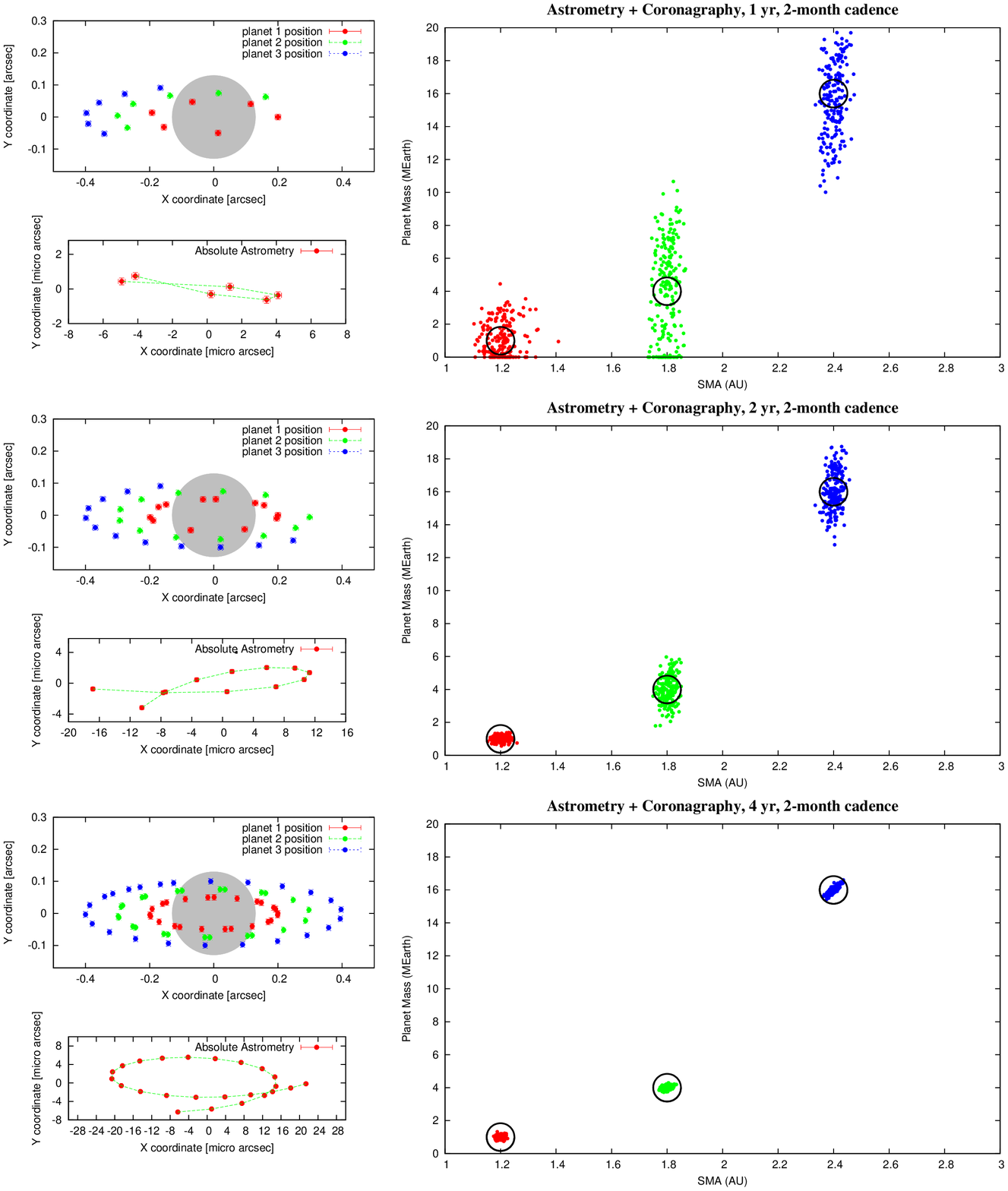} 
   \end{tabular}
   \end{center}
   \caption[multiplanetsol] 
   { \label{fig:multiplanetsolCA} 
Result of the astrometry + coronagraphy mission scenario, observing a three planet system around a Sun-like star at 6pc. Values to be measured are shown on the left with measurement error bars (1$\sigma$). Two hundred realizations of the observations are simulated, and the best three-planet solution is shown on the right for each simulation in the semi major axis ($x$ axis) -- mass ($y$ axis) plane. Total mission duration is increased from top to bottom, from 1 yr (top) to 4 yr (bottom). The gray disk indicates the blind area (within coronagraph inner working angle) for the planet-to-star relative astrometry measurements. The three dark circles indicate the actual configuration for the three-planet system.}
   \end{figure*}

A number of simplifications to the observing model have been used for the simulations.
\begin{itemize}
\item{The orbits are assumed to be circular and coplanar. This leaves a total of 12 free parameters: mass, SMA and orbital phase for each of the three planets, in addition to the stellar mass, its distance and the common orbital plane inclination.}
\item{The planets are assumed to recognizable in the coronagraphic images: provided that a planet is seen outside the coronagraph's IWA, it is assumed that it cannot be confused with other planets (this would be the case if each planet had recognizable colors)}
\item{The planets are detected in the coronagraphic images if and only if outside the coronagraph's IWA, and the precision of the planet's measured image position is independent of its location in the image or phase. In reality, the measured planet position would gradually become more uncertain as it is closer to the coronagraph's IWA, and contrast/sensitivity limits would prevent its detection in coronagraphic images when illumination fraction is small. This assumption leads to an underestimate of the required number of observation epochs.}
\end{itemize}
We use again an MC approach to determine how each mission scenario would perform on the simulated three-planet system. A total of 200 MC simulations were conducted in each case, using a normal distribution to produce each noisy data set. For each of the 200 MC simulations, a random initial solution is chosen with three planets as a starting point for a $\chi^2$ minima search, with initial masses randomly chosen in the $0.5 M_i$--$1.5 M_i$ range, where $M_i$ is the true mass of planet $i$, and with initial SMA randomly chosen in the $0.5 {SMA}_i$ -- $2.5 {SMA}_i$ range, where ${SMA}_i$ is the true SMA for planet $i$. Similarly, the initial orbital phase is randomly chosen within 1 rad of the true orbital phase. The sine of the planetary system inclination $\sin{incl}$ is randomly chosen with uniform probability distribution between 0 and 1. This choice of initial conditions ensures that the search for the optimal solution is allowed to find minima outside the true solution of they exist. A simulated annealing is used to find the solution with the smallest $\chi^2$, with 5e6 iterations. The estimated values for the 12 free parameters are derived from this maximum likelihood solution. 

Results of this simulation are shown in Figures \ref{fig:multiplanetsolA} and \ref{fig:multiplanetsolCA} for the astrometry-only and astrometry+coronagraphy mission scenarios. Measurements and solutions are shown for variable mission durations, and solutions are shown in the two-dimensional SMA versus planet mass space.

\subsection{Results}

Figure \ref{fig:multiplanetsolA} shows how well the planetary system characteristics are estimated for the astrometric mission, assuming a total mission duration ranging from 1 yr (top) to 4 yr (bottom). With a 1 yr long mission, none of the planets is unambiguously detected, and the measurements only show the existence of at least one planet with orbital period exceeding the mission duration. With a 2 yr mission duration, the astrometric data can exclude the presence of massive planets at small angular separation: no solution places a planet of more than 10 Earth mass within the central 1.5 AU. With a 4 yr long mission, the three planets are identified, but their masses and SMAs are poorly constrained (relative errors are approximately 10\% for the SMA and 30\% for the mass). There is some confusion between planets 1 and 2, and a correlated error between the mass and SMA for each of the planets. These results show that identification of the potentially habitable inner planet in this case requires about 4 yr.


Figure \ref{fig:multiplanetsolCA} shows how well the planets masses and SMAs are constrained for combined coronagraphy + astrometry missions with 1 yr, 2 yr and 4 yr durations. After only 1 yr, the three planets are clearly identified, even if for the two inner planets, there is no strictly positive lower limit to the planet mass (detection is then confirmed by the coronagraphic images). The inner planet is unambiguously identified as a low-mass ($< 5 M_{Earth}$) planet in the star's habitable zone. The simulation therefore demonstrates that a 1 yr mission duration (with measurements at only six epochs) is sufficient to identify high-priority targets for follow-up. If only astrometric measurements were used, it would take approximately 4 yr to reach the same conclusion. This figure also shows that the outer planet's mass is relatively well constrained even though it has only covered about one quarter of its orbit. With a 2 yr mission duration, the relative errors in the masses and SMA estimates are better than what is obtained in the astrometry only mission in 4 yr.

\subsection{Discussion}

Results shown in Figure \ref{fig:multiplanetsolCA} demonstrate that a combined astrometry + coronagraphy mission can be deployed with no a priori knowledge of the target locations, and quickly (in 1 -- 2 yr) identify potentially habitable planets. This is a significant advantage over the astrometry only case for which, as described in the Appendix, proper identification of planets requires that all planets of the system are either identified by RV or cover $>80\%$ of their orbit during the measurement span. This fast identification would allow the first phase of the an imaging mission + astrometry mission to perform high-cadence observations of a large number of targets, while the second phase of the mission could focus on acquiring high-quality spectra and astrometry of the low-mass planets identified in the habitable zone of the sample stars. 

Figure \ref{fig:multiplanetsolCA} illustrates that a key advantage of the combined measurement, beyond the improvement in mass and orbital parameters estimations, is the very low probability of obtaining a false solution when solving for the orbital elements and masses of multiple planets. With astrometry alone (Figure \ref{fig:multiplanetsolA}) or combined RV and astrometry measurements (see the Appendix), the probability of obtaining false solutions is non-negligible and rapidly increases as the measurement S/N or number of measurements decrease. The combined astrometry and coronagraphy measurement is very robust against such errors even for low astrometry S/N. This good performance in determining the masses and orbits of planets in multiple systems is particularly relevant in light of recent $Kepler Space Telescope$ results. \cite{2011ApJS..197....8L} find that 408 of the 1200 recently released $Kepler$ candidate planets are in multiple transiting systems. Given that the orbits of planets in these systems must be co-aligned to $\approx$ 1$^{\circ}$ to be detected by $Kepler$, this suggests that all or nearly all planets may be in multiple systems if the orbital inclinations of planets are aligned to only $\approx$ 10$^{\circ}$ in each system.


While the planets chosen for this example are expected to have similar reflected light contrasts (the increased planet mass as a function of angular separation results in roughly constant surface area divided by squared SMA), the astrometric signal is dominated by the outer planet (32 times stronger signal than the inner planet). The results presented in this section thus indicate that the relatively small astrometric signature of a habitable planet can be reliably separated from the much stronger signal of outer planets. The ability to identify and measure the mass of planets as a function of planet mass and separation is expected to follow the trends discussed in Section \ref{ssec:singleplanetdiscussion}: at the small separations, the coronagraph's IWA limits sensitivity, while the coronagraph contrast, and possibly mission duration, limit sensitivity for outer planets. With multiple planets, confusion between planets may also become an issue, and could be especially serious for a system viewed edge-on, for which the number of independent measurements is essentially halved. Planet colors and brightness as a function of illumination phase (ignored in this paper) would then need to be taken into account to help identify individual planets. We note that the proposed astrometry + coronagraphy measurement scheme is however well suited to this challenging configuration, compared to a much more degenerate astrometry only measurement scheme.

We note that our simulations do not assume availability of RV measurements, which have been shown to greatly mitigate confusion issues for an astrometry-only mission (see the Appendix). RV measurements are also expected to be valuable for an astrometry+coronagraphy mission, as massive distant planets would not be visible in reflected light in the coronagraphic images, but would produce large-amplitude long-period astrometric signals. Unknown planets within the coronagraph IWA will produce astrometric noise for the combined astrometry + coronagraphy approach described in this paper. These planets will have shorter period, and, at constant mass, a smaller astrometric signature than planets outside the IWA. They can therefore only compromise the solution if significantly more massive than the planets for which the mass is to be measured. This problem can be mitigated by a larger number of measurements, especially at high cadence, to (1) identify the inner planet(s) from astrometry alone and (2) average down the astrometric noise from unknown short-period planet(s). RV measurements, if available, would be very well suited to help identify such planets.

\section{Deep wide-field imaging sensitivity with a Diffractive Pupil Telescope}
\label{sec:deepim}

The effect of the dots on the telescope primary mirror on general-purpose wide-field imaging is evaluated in this section. The baseline design described in \cite{DPTpaper1} is adopted in this paper: a 1.4 m telescope with dots covering 1\% of the primary mirror area, with a 0.29 deg$^2$ field of view camera with 44 mas pixels observing in visible light. We assume in this section that this system is observing a Sun-like star at 6pc ($m_V = 3.7$).

\subsection{Light Lost to the Dots}
The light lost due to the dots is the sum of the light directly absorbed by the dots and the light they diffract out of the point spread function core into a wide halo of spikes. This second component is diffracted out to large distances, and is considered lost for science. Both quantities are equal to the fractional area of the pupil occupied by the dots. In the baseline design chosen, the total photometric loss is therefore 1\%+1\% = 2\%. We note that this loss is small compared to the sensitivity gain offered by adopting an unobstructed pupil, as the photometric sensitivity loss due to spiders and central obstruction in an on-axis telescope design would be larger than 2\%.

\subsection{Additional Background Due to Light of the Central Star Diffracted by the Dots}
The central star's diffraction spikes extend over most of the wide-field image. Although their contrast relative to the central star is faint (approximately $10^{-8}$ along the spikes), the central star is much brighter than other sources in the field.

Over 50\% of the field, the additional diffracted light is less than 2.8 photons per day per pixel. Table \ref{tab:scatlight} shows how, for a $m_V = 3.7$ central star, the light from the central star compares with the zodiacal light background. The 50 percentile line shows that the median diffracted light surface brightness is 0.03\% of the zodiacal light brightness. Over 95\% of the field, the additional light introduced by the dots on the primary mirror is less than 1\% of the zodiacal background. If the telescope were pointed on Sirius, 95\% of the field sensitivity would be photon-limited from zodiacal light, while over 5\% of the field, the photon noise from the spikes would limit sensitivity. The zodiacal light photon-noise-limited point source detection limit for the 1.4 m diameter telescope is $m_V \approx 32$ for a 2 day unfiltered exposure, and would be $m_V \approx 34$ for a full set of 32 observations, each 2 day long.

The diffraction spikes can also be numerically removed from the wide-field image to a high accuracy, as they are, to first order, static on the detector. Since the observation mode has the telescope slowly rolling during the observation, background sources are rotating on the detector while the spikes are static. A median of the images acquired during the roll contains only the spikes (and the zodiacal light background), and can then be subtracted to each individual frame prior to de-rotation and co-addition.

\begin{deluxetable}{ccc}
\tablecolumns{2} 
\tablewidth{0pc} 
\tablecaption{\label{tab:scatlight} Scattered Light Due to Primary Mirror Dots} 
\tablehead{ 
\colhead{Field Percentile\tablenotemark{a}} & \colhead{Flux($photon.s^{-1}.pixel^{-1}$)\tablenotemark{b}} & \colhead{Flux/Flux$_{zodi}$}\tablenotemark{c}}

\startdata 
1 \%      & 8.28 10$^{-6}$ & 7.66 10$^{-5}$\\
5 \%      & 1.08 10$^{-5}$ & 9.96 10$^{-5}$\\
10 \%     & 1.28 10$^{-5}$ & 1.19 10$^{-4}$\\
20 \%	  & 1.68 10$^{-5}$ & 1.55 10$^{-4}$\\
50 \%     & 3.23 10$^{-5}$ & 2.99 10$^{-4}$\\
80 \%     & 9,27 10$^{-5}$ & 8.56 10$^{-4}$\\
90 \%     & 3.06 10$^{-4}$ & 2.82 10$^{-3}$\\
95 \%	  & 1.19 10$^{-3}$ & 1.10 10$^{-2}$\\
99 \% 	  & 4.35 10$^{-2}$ & 4.02 10$^{-1}$
\enddata 
\tablenotetext{a}{All fluxes are measured in a 3$\times$3 arcmin box centered 6 arcmin from the optical axis. Scattered light becomes smaller at larger separations due to the Airy pattern contribution and the diffraction envelope of the spikes (equal to the diffraction pattern of a single dot).}
\tablenotetext{b}{All values in this table assume an unfiltered exposure (zero point = $5.5^{10} \: photon.sec^{-1}$), and a $m_V = 3.7$ central star}
\tablenotetext{c}{Assuming $m_V = 22.5 \: magn.arcsec^{-2}$ zodiacal light background level}
\end{deluxetable}


\section{Conclusions}

The DPT design allows simultaneous deep wide-field imaging, coronagraphic imaging, and astrometric mass determination of exoplanets around nearby stars. We have shown and quantified in both this paper and our previous publication that the measurements can be performed without impacting each other's performance: the coronagraphic and deep wide-field observations are not significantly affected by the presence of dots on the primary mirror. The potential cost and complexity saving offered by combining three separate missions into one observatory could be significant, and we are thus currently evaluating the feasibility and performance of this concept through both simulations and laboratory demonstrations.

We note that the benefits of performing simultaneously the three observations extend beyond the quantifiable metrics (planet mass and orbital parameters uncertainties) discussed in this paper. Together, astrometric measurements and coronagraphic images are a powerful way to avoid confusion problems in complex planetary systems, where a large number of planets, comets, dust clouds may be present. Astrometric measurements may also allow recovery of planets which are at the limit of detection in the coronagraphic images, for example because of residual structure in the scattered light halo: the astrometric measurements could confirm or eliminate features which could be either a planet or a speckle. This is particularly beneficial if the astrometric S/N is high but the contrast of the planet makes its identification in coronagraphic images ambiguous -- images acquired at different epochs can then be co-added following the planet location to reach deeper contrast (by the square root of the number of measurements) than otherwise possible. This scheme becomes especially valuable for planets at the outer edge of the habitable zone, which are fainter but have stronger astrometric signatures than planets closer in.

We note that these benefits do not require strict simultaneity of the astrometric and coronagraphic measurements as long as the two types of measurements can be connected together. Separate astrometry and direct imaging missions, if operating within a few years, would thus offer the same advantages, provided that the astrometry measurements are not performed after the direct imaging measurements (this would not allow prioritization of targets for follow-up spectroscopy of potentially habitable Earth-mass planets with the direct imaging mission).

Acquiring deep wide-field images around nearby stars may also reveal extended debris disks (analog to the Kuiper Belt, scattered disk, or Oort Cloud in the solar system), and therefore advancing our understanding of the planetary system architecture and history. Finally, we note that the spikes image in the wide-field camera are a low-resolution spectrum of the central star, which, when acquired simultaneously with the coronagraphic and astrometric measurements, may allow additional calibration. 

Our study also shows that coronagraphic imaging and astrometry allow stellar mass measurements to percent-level precision for any star around which planets can be identified, allowing calibration of the stellar mass-luminosity relationship.

A laboratory effort is under way to validate the concept and demonstrate that sub-$\mu$as level astrometric precision is attainable with a wide-field imaging telescope. A separate effort is aimed at laboratory demonstration that the DTP concept is compatible with high contrast coronagraphic imaging. Results from these experiments will be reported in future publications.

\acknowledgments
This work is funded by the NASA Astronomy and Physics Research and Analysis (APRA) program and the State of Arizona Technology Research Initiative Fund (TRIF). Support for this work was also provided by NASA through Hubble Fellowship grant HST-HF-51250.01-A awarded to S. Mark Ammons by the Space Telescope Science Institute, which is operated by the Association of Universities for Research in Astronomy, Inc., for NASA, under contract NAS 5-26555.

\bibliography{ms}

\appendix

\section{Finding Planets from astrometry only}
\label{ssec:astrometryonly}

In this Appendix, we briefly review what can be detected by astrometry alone, a scenario that has been studied as part of the Space Interferometry Mission (SIM) project. In addition to astrometry, it is here assumed that all the nearby stars will have been monitored by RV measurements. While RV cannot detect 1 $M_{earth}$ mass planets in a 1 yr orbit, due to stellar RV noise, RV because of its long time baseline will have found larger planets Jupiters and Neptunes with long orbital periods. A detailed study of what is possible when 5 yr of precision astrometric data plus $\approx$15 yr of 1 m.s$^{-1}$ RV data was conducted during a "double blind" simulation of a 5 yr SIM. We briefly summarize in this section the lessons learned during this double blind study. A more detailed description of the double blind study is provided in \cite{2010EAS....42..191T,2010ASPC..430..249T}.

The first conclusion is that astrometric detection of exo-Earths requires a mission S/N=6. Here S/N is defined as signal-to-noise where signal is the semi-major axis of the astrometric orbit and noise is the $1\sigma$ rms error of the whole 5 yr data set. If there were 100 measurements during the mission and 12 parameters are solved for (five stellar parameters: position, proper motion, and parallax; six orbital parameters and the mass of the planet), the $1\sigma$ 5 yr mission noise is the single epoch noise divided by $\sqrt{100-12}$.  With an S/N$>6$, the false alarm probability (FAP) is $<$ 1\%. FAP of 1\% means that if there were no planets, and 100 stars are searched, noise will mistakenly be identified as a planet once. Second, at S/N$=6$, the mass of a 1 $M_{earth}$ planet would have a $\approx$25\% $1\sigma$ error. The orbital phase in radians has an error that is $\approx$0.5 of the mass. The period is measured more accurately for short-period planets, a planet with a 2x shorter period have its period measured twice as accurately. Last of all, the ability to find a planet and measure its mass was independent of the planet's period as long as the period was less than $\approx$80\% of the mission length. The sensitivity degraded with longer periods, by about 2x when the planet period equaled the mission length.

The main purpose of the double blind test was to examine the astrometric detection of multiple planets. In general the presence of multiple planets does not present a problem for astrometric detection if the planet orbital frequencies are separated more than 1/mission length. The double blind test used a large variety of multiple planet systems and in general confusion in the frequency domain was not an issue except for planets whose periods were longer than the mission length. The parallax effect has a 1 yr period and planets whose orbital periods were between 0.9 and 1.1yr would have part of their motion absorbed into the parallax solution. Again out of a randomly selected 100 planetary systems only one had an earth like planet close enought to 1.0yr to present a problem but by chance its orbital plane was almost orthogonal to the parallax signature and its mass was properly measured. If a multiple planet system had one large outer planet, such as Jupiter in a 12 yr orbit, it was still possible to "fit" for the 12 yr planet and after that find all the planets with shorter periods whose S/N was $>$ 6. The problem arose when there were two outer planets, for example with 12 and 20 yr periods. In this situation fitting a single long-period planet to the data when there were actually two long-period planets could result in significant residuals. The fitting error in general is an "arc" of motion over the 5 yr data period. But because proper motion is also fitted at the same time, the arc becomes a closed curve, mimicking a planet with a period slightly less than the mission lifetime. With experience the teams participating in the double blind study were able to recognize this about 70\% of the time and were then able to recover both long-period planets and the shorter period planets. When discussing the ability of astrometry to detect a planet in a multiplanet system, there are two statistical metrics, one is called completeness, what fraction of the planets whose S/N$>6$ and had periods less than 5 yr were actually found and the second is confidence, what fraction of "claimed" discoveries were actual planets. The result of the double blind test was very encouraging in that both completeness and confidence were above 90\%.

\end{document}